# Direct measurement of quantum phase gradients in superfluid $^4$He flow

Yuki Sato, Aditya Joshi and Richard Packard

Physics Department, University of California, Berkeley, CA 94720

We report a new kind of experiment in which we generate a known superfluid velocity in a straight tube and directly determine the phase difference across the tube's ends using a superfluid matter wave interferometer. By so doing, we quantitatively verify the relation between the superfluid velocity and the phase gradient of the condensate macroscopic wave function. Within the systematic error of the measurement (~10%) we find $v_s = \hbar/m_4 \nabla \phi$.

The modern description of superfluidity is a melding of two conceptual frameworks[1]. The first concept, due to Landau[2], envisions a two-component system. An inviscid "super" component carries zero internal entropy and is described by density $\rho_s$ and velocity $v_s$. A normal component carries the entire liquid's entropy, and is described by density $\rho_n$ and velocity $v_n$. This theory successfully describes many thermo-hydrodynamic situations. The second concept, introduced by London[3] and expanded by Onsager, Feynman[4] and Anderson[5], relates the concept of superfluidity with the existence of a condensate in the many-body system described by a wavefunction $\Psi = |\Psi|e^{i\phi}$. The connection between the two approaches is made through the statement that the superfluid velocity $v_s$ is proportional to the wavefunction's phase gradient $\nabla \phi$. More precisely,

$$v_s = (\hbar/m_4)\nabla \phi, \qquad (1)$$

where $\hbar$ is Planck's constant ($h$) divided by $2\pi$ and $m_4$ is the $^4$He atomic mass. The main physical consequence of this connection is the quantization of superfluid circulation, which is involved in interpreting and understanding many experiments dealing with rotating helium,[6] turbulence[7] and the temperature dependence of superfluid persistent currents[8,9]. However, a direct measurement relating quantum phase difference and flow velocity has remained elusive for the lack of a phase measuring device. Using a superfluid $^4$He interferometer, we have now independently determined both $v_s$ and $\nabla \phi$ and quantitatively confirmed their relationship.

Our apparatus is schematically shown in Figure 1a. The topmost tube (of interior length $\ell = 2.5 \pm 0.05 cm$ and cross-sectional area $\sigma = (3.78 \pm 0.04) \times 10^{-2} cm^2$) contains a heater at one end. The opposite end of the tube terminates with a thin roughened copper sheet whose Kapitza boundary resistance dominates the thermal contact between the entire inner flow region and the surrounding superfluid helium bath. The apparatus is immersed in this bath whose temperature is maintained at a few millikelvins below the superfluid transition. We create uniform $v_s$ along the top tube and use a superfluid helium quantum interference device (SHeQUID)[10] to directly measure the corresponding phase difference $\Delta \phi$.

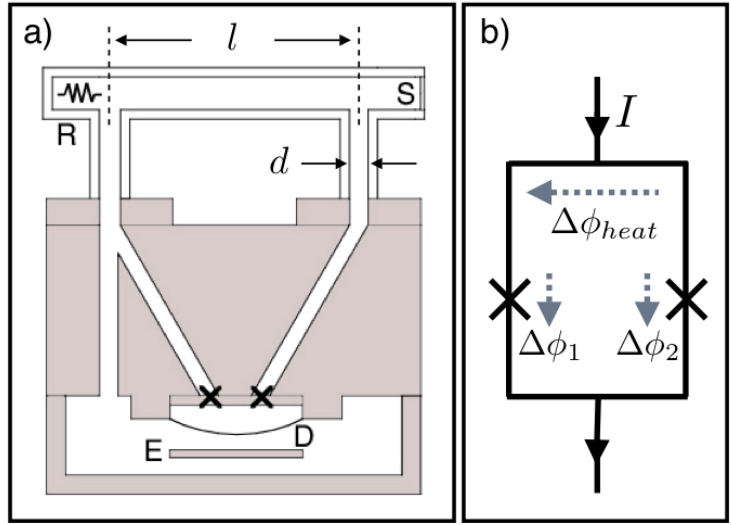

**Figure 1**. a) Experimental apparatus. The inside is filled with superfluid $^4$He and the entire apparatus is immersed in a bath of liquid helium. A resistive heater (R) and a thin Cu sheet (S) serve as a heat source and a temperature sink. The top tube and the two connecting arms are made of Stycast 1266 (insulating) to minimize the heat loss through the walls. Crosses indicate the aperture arrays. Each array consists of 100x100 30nm apertures spaced on a 3micron square lattice in a 60nm thick silicon nitride window. Flexible diaphragm (D) and electrode (E) form an electrostatic pressure pump. The diaphragm also forms the input element of a sensitive microphone based on superconducting electronics that are not shown. b) Equivalent SQUID circuit.

In the two fluid description[1], heat is carried by the normal component which flows away from the heater with velocity $v_n$ while the super component flows towards the heater with velocity $v_s$. Since there is no associated net mass current, it follows that $v_s = -\rho_n v_n / \rho_s$. The heat current $\dot{Q}$ is carried by the specific entropy (per unit mass) $s$, which resides entirely within the normal component. Thus $\dot{Q}/\sigma = \rho v_n T s$ and

$$|v_s| = \frac{\rho_n}{\rho \rho_s T s \sigma} \dot{Q}. \qquad (2)$$

Here $T$ is the temperature and $\rho$, $\rho_n$ and $\rho_s$ are the total, normal and superfluid densities respectively. From Eqs. (1) and (2), the phase gradient in the top tube should be

$$\nabla \phi_{heat} = \frac{m_4}{\hbar} \frac{\rho_n}{\rho \rho_s T s \sigma} \dot{Q}. \qquad (3)$$

As shown in Fig. 1a the top tube forms one arm of a superfluid interferometer, which contains two arrays of nanometer-sized apertures. Well below $T_\lambda$ the aperture arrays are characterized by a linear current-phase relation with discrete $2\pi$ phase slips. Closer to $T_\lambda$ they are described by a sine-like dc-Josephson current-phase relation[11,12,13]. In operation we apply a chemical potential difference $\Delta\mu$ (combining pressure and temperature differences) across this pair of aperture arrays. In response (in both the phase slip regime and the Josephson regime) each array exhibits mass current oscillations at a Josephson frequency $f_J = \Delta\mu/h$ that are detected by the microphone placed nearby).

We maintain the mass current oscillation frequency $f_J$ constant (typically near 700Hz) by a feedback technique. The combined oscillation amplitude $I_t$ from two arrays exhibits interference depending on the relative phase differences $\Delta\phi_1 - \Delta\phi_2$ that exists between them. For the SHeQUID the combined amplitude can be written as $I_t \propto |\cos[(\Delta\phi_1 - \Delta\phi_2)/2]|$ for arrays with equal oscillation amplitudes, and this quantum interference has been demonstrated[10] not only in a weakly-coupled Josephson regime but also well into a strongly-coupled phase slip regime.

When no currents flow in the interferometer there are no phase gradients and $\oint \vec{\nabla}\phi \cdot d\vec{l} = 0$ where the phase integral goes around the interferometer loop. For sufficiently low flow velocities (i.e. below the velocity to create quantum vortices) this phase integral condition is maintained even though a finite $\dot{Q}$ induces $v_s$ in the top tube. We can then write $\Delta\phi_{heat} + \Delta\phi_1 - \Delta\phi_2 = 0$ (see Fig. 1b) since the phase differences across the remaining segments of the loop are all negligible. Using this relation, the oscillation amplitude detected can be written as

$$I_t \propto \left|\cos(\frac{\Delta\phi_{heat}}{2})\right|. \qquad (4)$$

We can combine Eqs. (3) and (4) by writing the phase gradient in terms of the phase difference: $\Delta\phi_{heat} = \ell\,\nabla\phi_{heat}$. (We note that the length $\ell$ here is uncertain by the diameter ($d \approx 2.2mm$) of the tubes connecting the heat flow pipe with the aperture arrays.) This gives

$$I_t \propto \left|\cos\left(\pi\frac{m_4}{h}\left[\frac{\ell}{\sigma}\frac{\rho_n}{\rho\rho_s Ts}\right]\dot{Q}\right)\right|. \qquad (5)$$

Thus the existence of a uniform phase gradient associated with superfluid flow implies that the amplitude of the SHeQUID microphone should vary cosinusoidally with the heat input and, if Eq. (1) is quantitatively correct, the periodicity of the pattern is determined by known or measurable parameters.

The apertures in the two arrays are not identical due to the limitations of nanofabrication technology. Therefore the arrays have different oscillation amplitudes $I_{0,1}$ and $I_{0,2}$. For this case a more general (than Eq. 4) total mass current oscillation amplitude is $I_t \propto [\cos^2(\Delta\phi_{heat}/2) + \gamma^2 \sin^2(\Delta\phi_{heat}/2)]^{1/2}$ where the asymmetry parameter[14] $\gamma = (I_{0,1} - I_{0,2})/(I_{0,1} + I_{0,2})$. Then, even for destructive interference, the amplitude does not go to zero, which is useful for our feedback circuit that maintains the mass current oscillation frequency constant.

Fig. 2 is an example of a plot (at fixed temperature) of microphone amplitude as a function of heat input in the tube. The solid line is a fit using the general function above for two arrays with unequal critical currents. The excellent fit strikingly demonstrates that there is indeed a phase gradient across the tube that is linear in the heat-induced superfluid velocity: $\nabla\phi \propto v_s$. To demonstrate Eq. (1) quantitatively we need to determine the proportionality constant between $\nabla\phi$ and $v_s$.

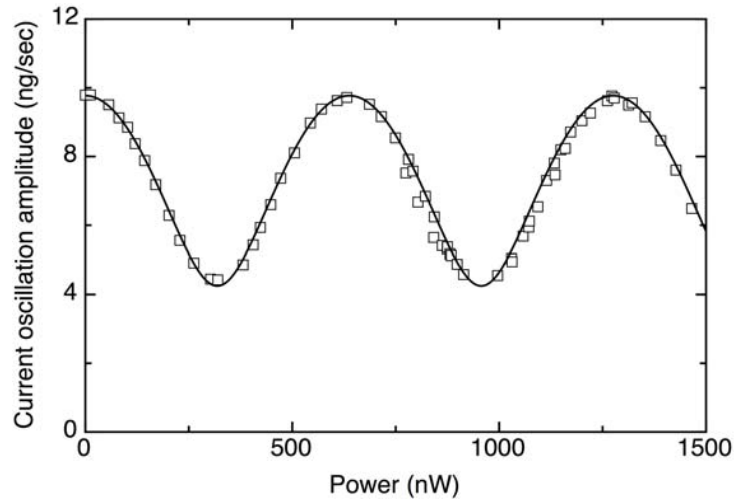

**Figure 2**. Measured current oscillation amplitude as a function of power put into the top tube. The solid line is a fit. These data are taken at $T_\lambda - T \approx 16 mK$. We have oriented our cryostat to catch just the right amount of rotation flux from the Earth in the interferometer loop so that the mass current oscillation amplitude is at maximum with zero power injected into the top tube[10]. A similar interference due to electron drift velocity has been seen in superconducting Josephson systems[15].

The heat current that leads to a $2\pi$ phase change across the tube can be seen from Eq. (5) to be

$$\dot{Q}_{2\pi} = \frac{h}{m_4}\beta(T), \qquad (6)$$

where $\beta(T) \equiv (\sigma/l)(\rho\rho_s Ts/\rho_n)$. Here, $\dot{Q}_{2\pi}$ is the distance on the horizontal axis of Fig. 2 between two adjacent maxima or minima. We display the measured $\dot{Q}_{2\pi}$ as a function of $T_\lambda - T$ in Figure 3. With published data[16] on $\rho_s$, $\rho_n$, $\rho$ and $s$; and the design values of tube length and cross section we have computed $\beta(T)$. We plot this function and multiply it by a constant to fit the data in Fig. 3. The best-fit multiplication factor is (9.1 ± 0.9) × $10^{-8}$ m$^2$/sec which agrees with the expected value (from Eq. 6) of $h/m_4 = 9.97 \times 10^{-8} m^2 /\sec$ within the systematic uncertainty (which is dominated by the effective length of the heat flow tube as described earlier).

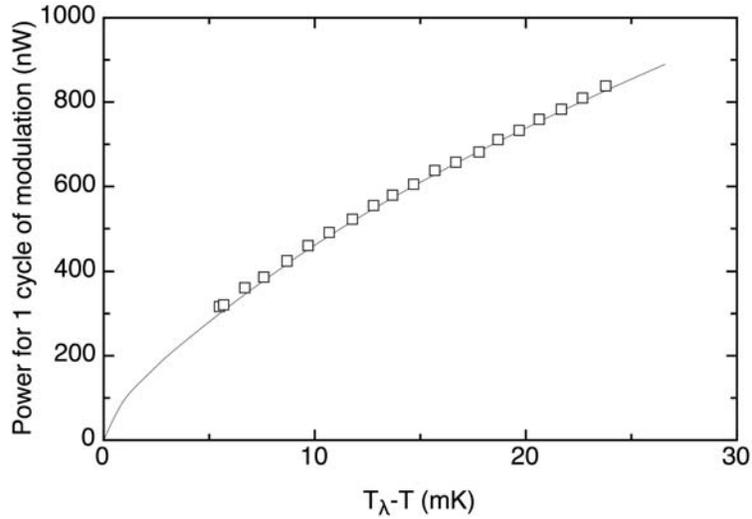

**Figure 3**. Power needed to cause the oscillation amplitude to move from one maximum to the next.

Figures 2 and 3 are the essential results of our experiment that directly demonstrates the fundamental relation linking quantum physics and the two fluid description of superfluid helium. It is clear that this superfluid paradigm is more than an idealist construct.

In this experiment we have also shown a method to electrically "inject" phase variations into a SHeQUID. This is a crucial element to develop a flux locked SHeQUID, analogous to techniques used for several decades in superconducting SQUIDs[17]. This important technical advance will permit the linearization of the intrinsically nonlinear interference relation underlying all interferometers.

**Acknowledgements**

We thank J. Treichler and D. Olynick for help with the fabrication of aperture arrays and E. Hoskinson for the early versions of the lab software as well as for sharing his insights on weak-link physics with one of us (Y.S.). We acknowledge helpful discussions with R. V. Duncan, D. Pekker, K. Penanen, and D. H. Lee. I. Hahn generously provided the high-resolution thermometer. J. Clarke suggested the term SHeQUID. This work was supported in part by NSF Grant No. DMR 0244882 and by the ONR. The aperture arrays were fabricated at the Cornell NanoScale Facility, a member of the NSF National Nanotechnology Infrastructure Network.

**Reference**

[1] D. R. Tilley and J. Tilley, *Superfluidity and Superconductivity* 3rd edn (Institute of Physics, Bristol and Philadelphia, 1990).
[2] L. D. Landau and E. M. Lifshitz, *Fluid Mechanics*, vol. 6 (Pergamon Press 2nd ed. July 1987).
[3] F. London, *Superfluids* (John Wiley & Sons, Inc., New York, 1950).
[4] R. P. Feynman, in Progress in Low Temperature Physics (ed. Gorter, C. J.) Vol. 1, Ch. II (North Holland, Amsterdam, 1956).

[5] P. W. Anderson, *Rev. Mod. Phys.* **38**, 298-310 (1966).

[6] R. J. Donnelly, *Quantized vortices in helium II* (Cambridge University Press, New York, 1991).

[7] C. F. Barenghi, R. J. Donnelly, and W. F. Vinen, *Quantized Vortex Dynamics and Superfluid Turbulence,* (Springer, Berlin, 2001).

[8] H. Kojima, W. Veith, E. Guyon, and I. Rudnick, *J. Low Temp. Phys*. **8**, 187-193 (1972).

[9] J. S. Langer and J. D. Reppy, in *Progress in Low Temperature Physics*, vol. 6, Chap. 1, ed. Gorter, C. J. (North-Holland, Amsterdam, 1970).

[10] E. Hoskinson, Y. Sato, and R. E. Packard, *Phys. Rev. B* **74**,100509(R) (2006).

[11] K. Sukhatme, Y. Mukharsky, T. Chui, and D. Pearson, *Nature* **411**, 280-283 (2001).

[12] E. Hoskinson, R. E. Packard, and T. M. Haard, *Nature* **433**, 376 (2005).

[13] E. Hoskinson, Y. Sato, I. Hahn, and R. E. Packard, *Nature Physics* **2**, 23-26 (2006).

[14] As in Ref. 10, we find that $|\gamma|$ is an increasing function of temperature very close to $T_\lambda$. This is expected because as the transition temperature is approached one of the $I_{0,1}$ will go to zero first thereby increasing $|\gamma|$ to its maximum value of 1.

[15] R. C. Jaklevic, J. Lambe, J. E. Mercereau, and A. H. Silver, *Phys. Rev.* **140**, A1628-A1637 (1965).

[16] R. J. Donnelly, R. A. Riegelmann, and C. F. Barenghi, The observed properties of liquid helium at the saturated vapor pressure. (University of Oregon, Eugene, Oregon, 1993).

[17] J. Clarke and A. I. Braginski, *The SQUID handbook: fundamentals and technology of SQUIDs and SQUID systems*, vol. 1 (Wiley-VCH, Weinheim, 2004).